# One- and two-dimensional correlation spectroscopy analyses for resonant and non-resonant coherent nonlinear optical processes

**SUPRIYA NAGPAL, BRYAN SEMON, GOMBOJAV O. ARIUNBOLD**[*]

*Department of Physics and Astronomy, Mississippi State University, Mississippi State, Mississippi 39769, USA*
*[*]ag2372@msstate.edu*

**Abstract:** Three-color coherent anti-Stokes Raman scattering represents non-degenerate four wave mixing process that includes both a non-resonant and resonant processes, the contributions of which depend on how the molecular vibrational modes are being excited by the input laser pulses. Non-degenerate four wave mixing processes are complex and understanding these processes requires rigorous data analytical tools, which still lack in this research field. In this work, we introduce one- and two-dimensional intensity-intensity correlation functions in terms of a new variable (e.g., probe pulse delay) and new perturbation parameter (e.g., probe pulse linewidth). In particular, diagonal projections are defined here as a tool to reduce both synchronous and asynchronous two-dimensional correlation spectroscopy analyses down to one-dimensional analysis, revealing valuable analytical information. Detailed analyses using the all Gaussian coherent Raman scattering closed-form solutions and the representative experimental data for resonant and non-resonant processes are presented and compared. This intensity-intensity correlation analytical tool holds a promising potential in resolving and visualizing resonant versus non-resonant four wave mixing processes for quantitative label-free species-specific nonlinear spectroscopy and microscopy.

## 1. Introduction

Correlation analyses have emerged as a powerful diagnostic tool in nuclear magnetic resonance, fluorescence, Raman, and infrared spectroscopies since its introduction in the early 1800s. Two-dimensional correlation spectroscopy (2DCOS) has been extensively applied to protein science, pharmacy, biomedical applications, and nano research. [1–3]. The main goal of 2DCOS investigations is to explore mutual correlations between perturbation-induced responses in spectroscopy. In the last five decades, significant progress has been demonstrated in developing two-dimensional correlation spectroscopy as an analytical tool to enhance and reveal the synchronously and asynchronously correlated features on 2D data maps. In the past, several techniques, such as moving window 2DCOS and null space projection, have advanced comprehensive interpretation of the results [4,5]. For example, by spreading the overlapped spectral features along the second dimension, apparent spectral resolution is also enhanced. Due to its application versatility, publication rates are about 180 articles per year, and almost 2.5K articles on 2DCOS have been published so far [6].

Over the last few years, the original 2DCOS tool [7] has been repeatedly updated according to the particular phenomenon under investigation. For example, two-trace two-dimensional correlation spectroscopy (2T2DCOS) was used by Yang et al. [8] to discriminate between pure sesame oil and it adulterated with corn oil. 2T2DCOS [9] uses only a pair of spectra instead of the whole sequential series. The authors introduced a diagonal projection approach [6] studying entire upper and lower diagonal sums [10,11], which will be extended to asynchronous 2DCOS, in this work. Another example is the systematic absence of cross peak (SACP) adopted by Guo et al. [12,13] used to extract pure component spectra from bilinear data obtained by hyphenated measurements, like chromatography-spectroscopy with elution time as the perturbation. Thomas and Richardson [14] introduced the community with the model of moving window 2DCOS, where the data is divided into small windows in the perturbation interval to be analyzed. This concept was further extended by Shigeaki Morita et al., who developed perturbation-correlation moving window 2D correlation analysis [15]. To obtain sequential information, Nishikawa et al. used a new 2D-IR correlation analysis class at a given deformation Fourier frequency [16]. Mixtures that contain multiple components or multiphase composite material fall into the category of systems where even 2D spectra sometimes become too congested. Techniques like the 2D correlation of 2D spectra known as quadrature spectra were demonstrated by Noda [17] applied to hetero-spectral or multiple mode analysis. In [18], Noda studied projection analysis in 2DCOS such as null space projection or the so-called positive projection technique, where a projecting vector is employed to filter out contributions that obscure other features of 2D spectra.

2DCOS has been extensively adopted to incoherent Raman scattering processes but rarely used to coherent Raman processes [11]. The authors applied the 2DCOS technique to the coherent anti-Stokes and Stokes Raman scattering (CARS/CSRS) spectroscopy for studies of noise correlation [10] and hydrogen bonding in pyridine-water mixtures [11]. CARS is the third order resonant nonlinear optical four wave mixing (FWM) process [19–21]. CARS spectroscopy has become a powerful spectroscopy and microscopy technique [21,22,23]. In three-color CARS is an example of non-degenerate FWM, which involves two excitation pulses (pump and Stokes) and a delayed probe pulse. The advantage of three-color CARS is a possibility in resolving the resonant from non-resonant FWM processes [24]. In recent work, the authors observed so-called deferred CARS buildup by studying how the signal gradually builds up as a probe pulse is delayed with respect to pump and Stokes arrivals. This effect results from resonant excitation of Raman vibrational ring modes for the selected substance [24]. When tuned to far off-resonant excitation of Raman vibrational modes, the signal showed no deferred signal buildup effect. Hence, the deferred signal buildup is a resonant phenomenon in nature, which also depends on the Raman line width, widths of input pulses [25,26]. However, this effect can often be hidden under overwhelming non-resonant FWM signals. Thus, revealing the resonant signal requires more rigorous data analytical tools, which still lack in this research field. To address this challenge, in the current work, the authors introduce one- and two-dimensional correlation spectroscopy (1D/2DCOS) analyses as functions of probe time delay using probe width as an additional perturbative parameter.

## 2. All-Gaussian analytical approach for resonant versus non-resonant nonlinear optical processes

Exact closed-form solutions for non-resonant $P_{NR}(\omega,\tau)$ and resonant $P_R(\omega,\tau)$ responses for nonlinear four wave mixing processes are given by [24,25]

$$P_{NR}(\omega,\tau,\Delta\omega_{pr}) = c_0 E_p E_S^* E_{pr} \frac{\Delta\omega_p \, \Delta\omega_s \, \Delta\omega_{pr}}{W} \times \exp\left(-\frac{\tau^2}{2t_{NR}^2} - \frac{2\ln2\delta^2}{W^2} + i\frac{\tau\delta\Delta\omega_{pr}^2}{W^2}\right) \quad (1)$$

$$P_R(\omega,\tau,\Delta\omega_{pr}) = P_{NR}(\omega,\tau,\Delta\omega_{pr})\frac{(-i)}{\Delta\omega_{NR}}\sum_{j=1}^{N} c_j \, F(\zeta_j) \quad (2)$$

The parameters in the equations above include probe time delay – $\tau$; signal detuning – $\delta$; signal frequency – $\omega$; field amplitudes of the input Gaussian pulses – $E_p$ (pump), $E_S$ (Stokes) and $E_{pr}$ (probe); constant coefficients – $c_0$ and $c_j$; spectral full widths at half maxima (FWHMs) - $\Delta\omega_S$ (pump), $\Delta\omega_S$ (Stokes), and $\Delta\omega_{pr}$ (probe); width parameter – $W^2 = \Delta\omega_p^2 +$

$\Delta\omega_S^2 + \Delta\omega_{pr}^2$ ; effective width of the non-resonant (NR) response – $\Delta\omega_{NR}^2 = (\Delta\omega_p^2 + \Delta\omega_s^2)\Delta\omega_{pr}^2 / W^2$; and effective time duration of the NR response – $t_{NR} = \sqrt{4\ln2} / \Delta\omega_{NR}$. The solution of Eq. (1) for NR process remains Gaussian with all-Gaussian input pulses. It determines the NR process, the intensity of which is given by $I(\omega, \tau, \Delta\omega_{pr}) = |P_{NR}(\omega, \tau, \Delta\omega_{pr})|^2$. Resonant nonlinear optical response given in Eq. (2) and its intensity $I(\omega, \tau, \Delta\omega_{pr}) = |P_R(\omega, \tau, \Delta\omega_{pr})|^2$ consists of vibrational modes (Raman lines) expressed by the so-called Faddeeva function [27–29]. The Faddeeva function $F(\zeta_j)$ is an error function with a complex argument $\zeta_j = [(\delta\Delta\omega_{NR}^2/\Delta\omega_{pr}^2 - \Delta_j + i\Gamma_j)t_{NR} - i\tau/t_{NR}]/\sqrt{2}$ where $\Gamma_j$ are the FWHM of the $j^{th}$ Raman line and $\Delta_j$ is the Raman detuning. The analytical expression for the time delay for maximum buildup signal is presented in [26,27] and later validated with the experimental observations in [25]. In this section, we consider pure NR and pure resonant processes only.

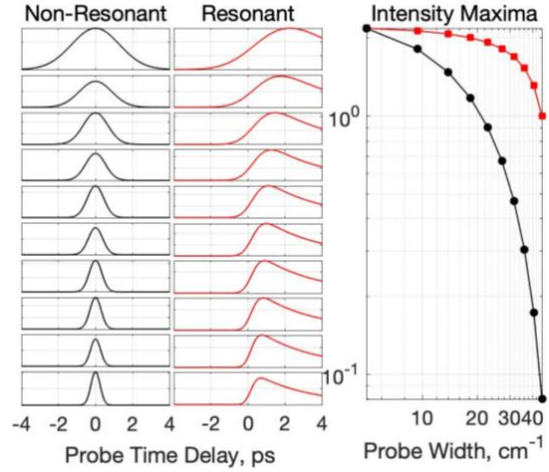

Figure 1: Intensities calculated from Eq. (1) (first column) and Eq. (2) (middle column) as functions of probe time delay. Intensity maxima for resonant (red curve) and non-resonant (black curve) processes in log-log scale (last column).

Figure 1 shows a spectral contrast in the pure resonant and NR responses and their dependence on the probe bandwidth and its temporal delay on the spectrally resolved profiles. The results in the first column of Fig. 1 are obtained using the NR spectral intensities $|P_{NR}(\omega, \tau, \Delta\omega_{pr})|^2$) at different spectral bandwidth of the probe. The pure resonant signals for varying probe widths are plotted in the middle column of Fig. 1 using $|P_R(\omega, \tau, \Delta\omega_{pr})|^2$. For resonant signals, an asymmetric shift in the peak position is observed as probe width gets narrow, but that is not the case for NR signals. The resonant responses also become maximal at some positive probe delay. The maximum signal gradually decreases with a decay rate determined by the dephasing time of molecular oscillations rather than the probe pulse profile. An all-Gaussian pulse model results in the output signal that is also a Gaussian. From the first column of Fig. 1, FWHMs for the non-resonant signal intensities are 3 ps for probe width <10 cm$^{-1}$ and 0.6 ps for probe width 45 cm$^{-1}$, which shows that width in temporal FWHM is decreasing as probe pulse width increases. From Fig. 1 (middle column), the resonant signal is peaked at 0.7 ps for the broadest probe width (45 cm$^{-1}$), but as the probe width gets narrower, the largest shift for the maximum signal is found to be 2.2 ps for <10 cm$^{-1}$ probe width. The shift in the resonant signal is coined the term deferred buildup time [24].

The last column in Fig. 1 is a log-log plot displaying the intensity change difference in the local maxima of the isolated resonant and NR responses to change probe width. The NR signal

(see black curve) shows a faster falloff than the resonant signal (see red curve). The maximum difference in the resonant and NR intensities for the probe width range shown here can be more than an order. Hence, we conclude that the best contrast between the two contributions into the overall integrated nonlinear response and the suppression of the NR background without a noticeable effect on the resonant signal's peak intensity can be controlled externally by the input laser pulse width variations.

## 3. One- and two-dimensional correlation spectroscopy (1D/2DCOS) analyses

In traditional 2DCOS [7,30], a series of spectra is obtained, where each one of them is measured under the influence of a discretely varying external perturbation. External perturbation (e.g., time, temperature, noise, or concentration) induces changes in the observed spectral intensities, correlation of which is studied on 2D maps. In this work, we define synchronous $\Phi(x_i, x_j)$ and asynchronous $\Psi(x_i, x_j)$ functions [7] as

$$\Phi(x_i, x_j) = \frac{1}{M} \sum_{k=1}^{M} I(x_i, y_k) \cdot I(x_j, y_k) \qquad (3)$$

$$\Psi(x_i, x_j) = \frac{1}{M} \sum_{k=1}^{M} I(x_i, y_k) \cdot \tilde{I}(x_j, y_k) \qquad (4)$$

However, here two independent variables $x_{i,j}$ stand for probe pulse delay $\tau$ with respect to arrival of the two excitation (pump and Stokes) pulses (but not frequency as in traditional 2DCOS), and the perturbation parameter $y_i$ stands for probe pulse width $\Delta\omega_{pr}$. We select a particular frequency $\omega$ in the signal spectrum and keep it constant in all correlation analyses. The Hilbert-Noda transformation of intensity $I(x_i, y_k)$ is given by

$$\tilde{I}(x_i, y_k) = \frac{1}{M} \sum_{l=1}^{M} N_{kl} \cdot I(x_i, y_l) \qquad (5)$$

where the Hilbert-Noda matrix elements are given by [31]

$$N_{kl} = \begin{cases} 0 & l = k \\ \frac{1}{\pi(l-k)} & l \neq k \end{cases} \qquad (6)$$

In general, Hilbert operation on a function produces another function of a real variable [32]. Traditionally, it can be explained as the following example when the Hilbert transform is represented in the frequency domain; it imparts phase-shift of $\pi/2$ to every Fourier component. An application of the Hilbert transformation twice results in the negative and positive frequency components getting shifted equivalent amounts equivalent to negative of the original function [33]. Thus, the repeated Hilbert-Noda transformation of $\tilde{I}(x_i, y_k)$ restores the original signal $I(x_i, y_i)$ as

$$I_{restored}(x_i, y_k) = -\frac{1}{M} \sum_{l=1}^{M} N_{kl} \cdot \tilde{I}(x_i, y_l) \qquad (7)$$

2D plots shown in Fig. 2 (a and d) are the normalized intensities as functions of both probe width change and probe pulse delay time change (scaled in picosecond - ps). Theoretical data for the NR and resonant processes by using the exact closed-form solutions given by Eqs. (1) and (2) are seen in the upper and lower row, respectively. The first, middle, and last columns show the normalized signal intensities, their single Hilbert-Noda transform (using Eqs. (5) and (6)), and the negative of their double Hilbert-Noda transforms(using Eq. (7)). The normalized intensity in Fig. 2(a) attributed to NR contributions can be seen to become symmetrically widened as the probe pulse duration increases (i.e., width decreases). On the contrary, the FWHM intensity in the resonant process (see, Fig. 2(d)) is no longer symmetrically increasing in width. As mentioned above, this asymmetric increase is a signature of the signal buildup effect where the maximum signal is delayed depending on the probe spectral width. The middle column in Fig. 2 for the Hilbert-Noda transforms, demonstrates symmetric and asymmetric changes. Symmetrically separated positive peaks are shown in Fig. 2(b) for NR processes. Asymmetric changes of positive peaks are shown in the case of resonant processes (see, Fig. 2(e)). The last column displaying the negative value of double Hilbert-Noda transforms, demonstrate a shape preserved and consistent restored intensity results compared to the original (before transforms) in the first column.

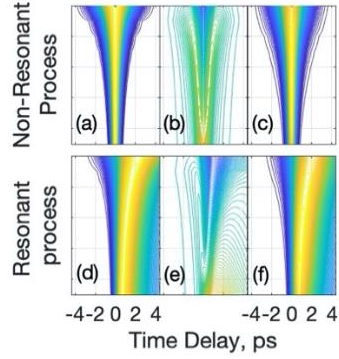

Figure 2: Normalized intensities (first column (a,d)) for and the results after single (middle column (b,e)) and double (last column (c,f)) Hilbert-Noda transformation of pure non-resonant (upper row) and resonant (bottom row) processes as functions of probe time delay.

In our previous work [10], we defined a one-dimensional (1D) correlation function in frequency domain results of which could directly be derived from the sum of the diagonal elements in the definition of the 2D synchronous correlation spectrum. The features of 2DCOS, such as the correlation peaks, were depicted as side peaks in the 1DCOS. In this work, we generalize the formula used in [10,11] to Eq. (8 and 9) for variables different from frequencies. We introduce the diagonal projection (i.e., diagonal directional sum) of the synchronous and asynchronous 2DCOS functions (see, Eq. (3 and 4)) in the following 1D forms as

$$G_\Phi^{(2)}(\Delta x_j) = \sum_{i=1} \Phi(x_i, x_i - \Delta x_j) = \frac{1}{M}\sum_{i=1}\sum_{k=1}^{M} I(x_i, y_k) \cdot I(x_i - \Delta x_j, y_k) \qquad (8)$$

$$G_\Psi^{(2)}(\Delta x_j) = \sum_{i=1} \Psi(x_i, x_i - \Delta x_j) = \frac{1}{M}\sum_{i=1}\sum_{k=1}^{M} I(x_i, y_k) \cdot \tilde{I}(x_i - \Delta x_j, y_k) \qquad (9)$$

where $G_\Phi^{(2)}(\Delta x_j)$ and $G_\Psi^{(2)}(\Delta x_j)$ denote the 1D second order (i.e., intensity-intensity) correlation functions. The theoretical simulations for synchronous 2DCOS $\Phi(x_i, x_j)$ (first column), asynchronous 2DCOS $\Psi(x_i, x_j)$ (middle column), and synchronous and asynchronous correlations 1DCOS $G_{\Phi,\Psi}^{(2)}(\Delta x_j)$ (last column) functions, defined in Eqs. (3), (4), (8), and (9), respectively. The autocorrelations (top row and middle row) and cross-correlations (last row) for NR and resonant intensities are depicted in Fig. 3. As mentioned above, probe time delay is the variable here, and probe width is the perturbative parameter. Vertical and horizontal axes $x_{i,j}$ for 2DCOS are, therefore, independent time delays of probe pulse, and the horizontal axis $\Delta x_j$ for 1DCOS is the difference of these delays. The synchronous 2DCOS plots show the simultaneous (i.e., coincidental) changes found in the signals' intensity as the probe width changes. NR signals exhibit a single Gaussian peak centered at the center (0,0) on the self-correlation (i.e., main diagonal) line in the autocorrelation synchronous 2DCOS map (see, Fig. 3(a)). The FWHM of this Gaussian function is about 1 ps, which is also the time duration of the overlap region of intensities with changing probe widths (see column one of Fig. 1).

Resonant signals, on the other hand, show an off-center (0.3,0.3) and asymmetric peak in the positive quadrant of the synchronous 2DCOS autocorrelation map (see, Fig 3(d)). The FWHM of this peak is around 3 ps, matching with the lower limit of the overlap time delay region of the intensities in the second column of Fig. 1.

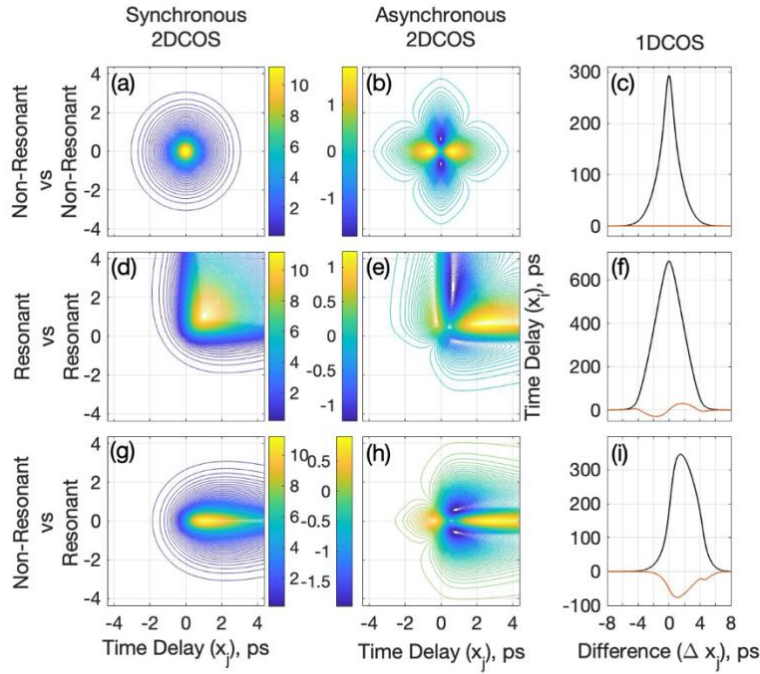

Figure 3: Synchronous 2DCOS (a-c), asynchronous 2DCOS (d-f) and 1DCOS (g-i: red for asynchronous and black for asynchronous) with autocorrelations of non-resonant and resonant processes (first and middle row) and cross-correlations (last row).

In synchronous cross-correlation 2DCOS map (see, Fig. 3(g)), the vertical and horizontal and out-of-page axes are devoted to the time delay of the NR, and resonant signals and the z-axis shows the intensities correlation of the two. When viewed along the horizontal axis, asymmetry can be seen in the peak's elongation as expected for resonant signals (FWHM of this peak is 3 ps). The asynchronous 2DCOS plots (middle column) exhibit the sequential or phase-related information between the two intensities being correlated. The asynchronous 2D autocorrelation of NR intensity (see, Fig. 3(b)) shows a symmetric well-known butterfly shape about the main diagonal with two positive local maxima (0.6,0.6) and two negative local maxima (-0.6, -0.6), both with FWHM of 1.2 ps and identical heights. The positive peak positions here give us the average change in the intensity of the NR signals (first column in Fig. 1). As expected, the autocorrelation resonant contribution is distorted in the symmetry, exhibiting two peaks in the first quadrant (positive and negative peaks at 1.6 ps), both having a FWHM of ~ 3.8 ps (in horizontal and vertical directions) corresponding to the higher end value of average FWHM of the resonant intensities in the middle column of Fig. 1, for these responses. The resonant signal's average shift in Fig. 1 (middle column) is found to be 1.2 ps. At this point, the intensities overlap; however, the positions 1.6 ps for the negative and positive peaks, mentioned above, correspond to the maximum change in the resonant responses. With an offset of 0.3 ps, the asynchronous 2DCOS cross-correlation (see, Fig. 3(h)) for NR versus resonant processes displays one peak positioned at 1.8 ps in the horizontal direction, with no contribution in the vertical direction (corresponding to resonant temporal delays) and FWHM of 3.4 ps matching the position and width of autocorrelation peak. In the last column of Fig. 3, the 1D correlation function has previously been employed [10,11] to show the same results as obtained from the diagonal projection (i.e., diagonal directional sum) of the synchronous 2DCOS (see black curves). Here we introduce a new definition of the correlation function for asynchronous 2DCOS (see red curves) as well. It is important to note that the second-order correlation function results obtained from Eq. (8) and (9) and the sum of elements of synchronous and asynchronous 2DCOS spectra along the on and off diagonals are identical. The horizontal axis in the 1DCOS plots is the change in the difference in the probe time delays. For example, a difference between -4 and 4 ps (or 4 and -4 ps) is reflected as -8 ps (or 8 ps). It should also be noted here that although the synchronous 1DCOS data are positive, the asynchronous 1DCOS data can be both positive and negative. Figure 3(c) shows the synchronous autocorrelation of NR intensities. That is a narrow, symmetric curve about zero delay (with FWHM of 2.2 ps). The synchronous autocorrelation for resonant intensities is plotted in Fig. 3(f). Its peak tends to shift, and the width becomes more expansive (with FWHM of 4.4 ps), see the black curve in Fig. 3(f). The asynchronous 1D autocorrelation function (see the curve in Fig. 3 (c)) for NR signals is flat and zero. This result is also reflected in the diagonal directional sum of the symmetric butterfly shape in the asynchronous 2DCOS plot, where positive and negative peaks cancel each other in the upper half (and also in the lower half) of the main diagonal (see, Fig. 1(b)). The asynchronous resonant 1D autocorrelation, on the other hand, exhibits two peaks with opposite signs (see the red curve in Fig. 3(h)), 1.7 ps about zero delay, both having a FWHM of 4 ps. The synchronous 1DCOS (see the black curve in Fig. 3(i)) exhibits a peak that is not only off-center (positioned at 0.7 ps) but is also asymmetric about the center. The FWHM of this curve is found to be 4.2 ps. It is made of unequal contributions of the rise and falling edges (maximum contribution from the falling edge showing positive delay). This maxima at half-width correspond to 2 ps delay in the buildup of resonant signals. The asynchronous 1DCOS depicts a negative peak at 0.3 ps with a FWHM of 3 ps. From the parameters obtained from the plots discussed above, the non-resonant responses clearly can be differentiated from the resonant responses. Therefore, this analysis provides an efficient approach to identify features in the 2D plots (synchronous and asynchronous) using a 1D plot.

## 4. Applications of 1D/2DCOS to three-color coherent anti-Stokes Raman spectroscopy (CARS)

In the previous sections, we studied pure resonant vs. NR responses. Both NR and resonant processes contribute to the CARS signal intensity as

$$I(\omega, \tau, \Delta\omega_{pr}) = |P_{NR}(\omega, \tau, \Delta\omega_{pr}) + P_R(\omega, \tau, \Delta\omega_{pr})|^2 \qquad (10)$$

In this section, we demonstrate the application of 1D/2DCOS analyses to three color CARS spectroscopy. The details pertaining to the experimental arrangement have been reported in [24].

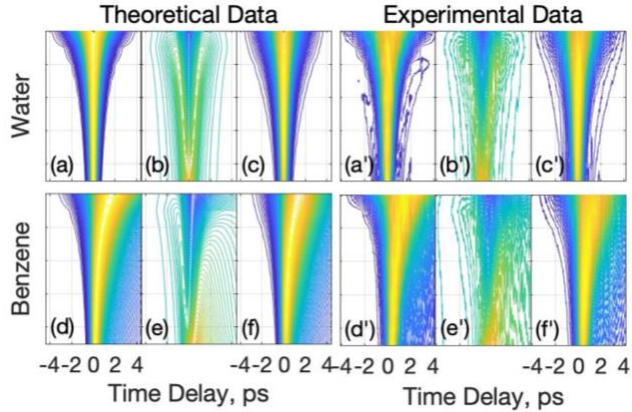

Figure 4: Theoretical (first three columns) and experimental data (last three columns) for intensity vs time delay 2D plots of water (upper row) and benzene (lower row). First column (a,d and a', d) show the normalized intensity with their Hilbert-Noda transform in middle column (b,e and b',e') and negative of their double Hilbert-Noda transform in last column (c,f and c',f')

Water is considered a sample, as it exhibits no Raman peak in our region of interest, thereby making it an ideal candidate to represent NR processes. On the other hand, two Raman active molecular compounds (benzene and pyridine) are chosen to visualize the resonant processes with minimal contribution from NR processes Further, to experimentally visualize the buildup delay time in coherent processes through 1D/2DCOS, only the transient changes of the spectra's selected wavenumber are considered. Keeping the frequency fixed, intensity correlations with the change in probe time delay are observed for benzene, pyridine, and water around 1000 cm$^{-1}$. In the benzene and pyridine spectra, the fixed frequency 994 cm$^{-1}$ corresponds to the breathing ring mode of molecular vibrations. The theoretical CARS signal is simulated, combining Eq. (1) and (2). The same as in Fig. (2), we plot the normalized intensities (in the first column), their single Hilbert-Noda transforms (using Eq. (5)) (in the second column) and the restored intensities by applying double Hilbert-Noda transforms as in Eq. (7) (in the third column). Results as a function of probe time delay for water (see, Fig. 4(a-c)) and benzene (see, Fig. 4(d – f)) are plotted in the upper and lower rows, respectively. To resemble the water sample, we keep only the NR contribution, whereas, for the benzene sample, we use only one resonant peak at 994 cm$^{-1}$ in the simulations. The theoretical results in Fig. 4(a-f) and experimental results in Fig. 4(a'-f') depict great settlement. Moreover, these results are similar to those shown in Fig. 2 for the resonant and NR processes. Hence, we conclude that the CARS process in the water (or benzene) sample strongly resembles the NR (or resonant) process. The normalized intensities for water in Fig. 4(a) can be seen to symmetrically widen for changing probe pulse spectral width about the zero delay; however, the normalized intensities for benzene (in Fig. 4(d)) depict an off-center (maximum found at 1.7 ps probe pulse delay) widening. This asymmetric widening is not only depicted in the intensities alone, but their Hilbert-Noda transforms (see, Fig. 4(b) and (e)) as well.

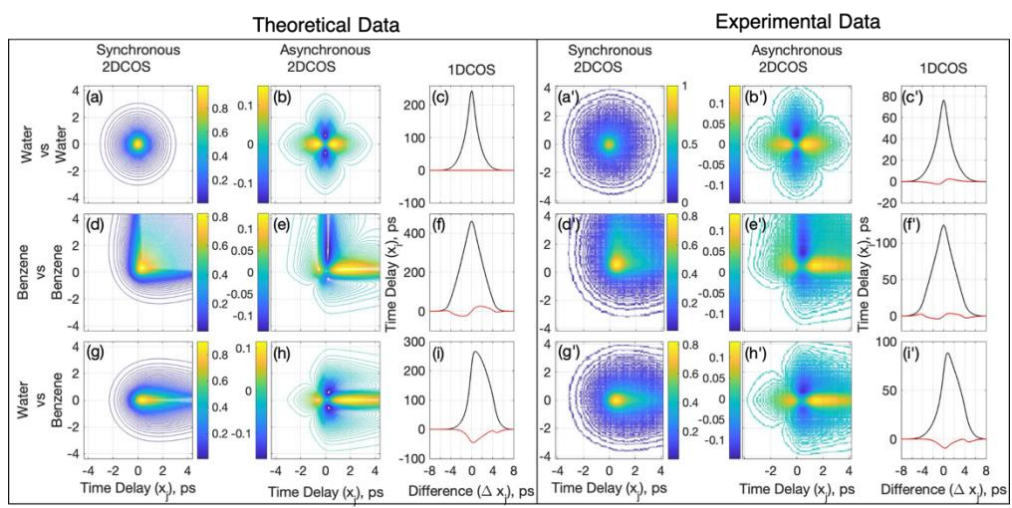

Figure 5: Theoretical (left set) and experimental (right set) synchronous 2DCOS (first column), asynchronous 2DCOS (middle column) and synchronous (black) and asynchronous (red) 1DCOS with autocorrelations of water and benzene processes (first and middle row) and cross-correlations (last row).

The first three columns (or the left set) in Fig. 5 (a-i) depict the theoretical simulations of 2D and 1D correlation results for water and benzene. The first and the middle rows are dedicated to the 1D/2D autocorrelations of water and benzene, respectively, and the last row shows the results for their cross-correlations. The experimental results are plotted in the same manner in the last three columns of Fig.5 (a'-i'). Water containing only NR contribution exhibits a single peak in the synchronous 2D autocorrelation plot (see, Fig. 5(a)) centered at zero of the probe pulse time delay. Most interesting results, however, are displayed in the cross-correlation of water and benzene (see, Fig. 5(c)). The peak is symmetric about the vertical axis, thus screening the water contribution and showing a peak shift in the horizontal direction, which screens the resonant process's contribution in benzene. This depicts the buildup delay time for the CARS signal to reach maximum intensity for benzene. In the resonant vs. NR 2D correlations discussed for Fig. (3), for the asynchronous autocorrelation 2D map for NR processes (see, Fig. 3(b)), a symmetric butterfly shape is observed that reappears in the asynchronous 2D correlation plot of water (see Fig. 5 (b)). The last three columns (or the right set) in Fig. 5 (a'-i') depict the experimental results of 1D/2DCOS results for water and benzene. The last columns in the left and right sets in Fig. 5 are devoted to the theoretical and experimental 1DCOS functions, respectively. The black curves represent the synchronous 1DCOS, and red curves represent asynchronous 1DCOS. The 1D autocorrelations for water can be seen in Fig. 5(c) and (c') for theoretical and experimental data, respectively. The single Gaussian peak in the synchronous 1DCOS autocorrelation of water is centered about the zero and has a FWHM of 2.4 ps. Despite considering only NR contributions from water, the 1D asynchronous water/water autocorrelation plot of water tends to show some little asymmetry in experimental results as compared to theoretical results (see red curves in Fig. 5(c) and 5(c')). This little asymmetry can be attributed to the fact that in the experiment, water is not a pure non-resonant responder but still contains far-off resonant molecules contributing a little bit. The 1D synchronous and asynchronous benzene/benzene autocorrelations can be seen in Fig. 5(f). The black curve depicting the synchronous correlation is found to have a FWHM of 4.2 ps in the difference of time delays. This change is twice that of a 2.1 peak shift in the time delay in asynchronous 2DCOS.

Next, to increase the complexity of the analysis, pyridine samples are considered, which contain two Raman vibrational resonant peaks in our region of interest. For the 1D/2DCOS analyses, two wavenumbers are chosen to be fixed, first being the ring breathing mode of pyridine (~994 $cm^{-1}$) and second being a wavenumber in the middle (~1014 $cm^{-1}$) of the aforementioned and ring bending mode of pyridine (1033 $cm^{-1}$).

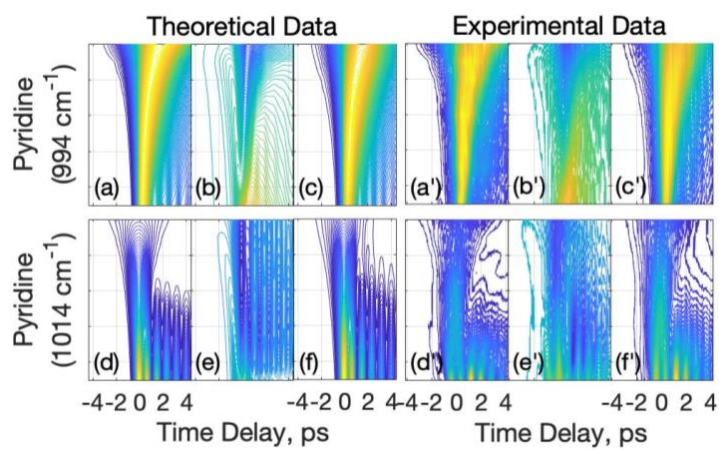

Figure 6: Theoretical (first three columns) and experimental data (last three columns) for intensity vs time delay 2D plots of pyridine at 994 $cm^{-1}$ (upper row) and at 1014 $cm^{-1}$ (lower row). First column (a,d and a',d) show the normalized intensity with their Hilbert-Noda transform in middle column (b,e and b',e') and negative of their double Hilbert-Noda transform in last column (c,f and c',f').

The theoretical CARS signal is simulated using Eq. (1) and (2) and is used to plot the normalized intensities (in the first column), their Hilbert-Noda transform (using Eq. (5)) in the second column, and the restored intensities (applying double Hilbert-Noda transforms as in Eq. (7)) in the third column respectively. Results as a function of probe time delay for pyridine (~994 cm$^{-1}$), (see, Fig. 6(a-c)) and pyridine (~1014 cm$^{-1}$) (see, Fig. 6(d–f)) are plotted in the upper and lower row respectively. The first three columns (or left set) show the theoretical simulations, and the last three columns (or the right set) show their corresponding experimental validations. It should be noted that results shown in the upper row are attributed to pyridine's vibrational mode; however, the results in the lower are due to the interaction of the two modes of pyridine and the incoming fields itself. Since there is no actual Raman peak at the second wavenumber considered, it can be said that the results in the latter case might resemble the results of NR processes. The plot features seen in the upper row of Fig. 6 can readily be compared to the plots in the upper row of Fig. 4 for benzene's resonant contribution. The reason for this similar behavior can be credited to the resemblance of the vibrational mode of pyridine and benzene at 991 cm$^{-1}$. Therefore, as compared to the normalized intensity vs. probe time delay 2D plot of pyridine (1014 cm$^{-1}$), the FWHM of pyridine (994 cm$^{-1}$) peak is asymmetrically widening about zero delay (see, Fig. 6 (a) and (d)). The 994 cm$^{-1}$ and 1014 cm$^{-1}$ peaks of pyridine, at the maximum and minimum probe widths, have a FWHM of 0.58 ps and 4.55 ps, respectively. When the probe spectral width is wide enough so that both pyridine modes are enclosed, beating between the two modes is observed. This phenomenon can explicitly be seen in the lower row of the intensity plots (see, Fig 6 (d -f)). A beating period of 0.87 ps is found from Fig. 6(d), which matches the wavenumber separation between pyridine's two vibrational modes (~38 cm$^{-1}$). The trace of beating can also be seen in one of the probe widths considered for pyridine (994 cm$^{-1}$) (see, Fig. 6(a-c)). When the probe width is very wide, some effect of beating can be seen in the pyridine (994 cm$^{-1}$) intensity plots as well (see Fig. 6 (a–c)). In Fig. 6 (a), the maximum response for the narrowest probe width lies around 2.53 ps and 0.69 ps for the widest probe width. This shift or buildup delay is consistent, as seen in the previous sections for benzene's resonant response.

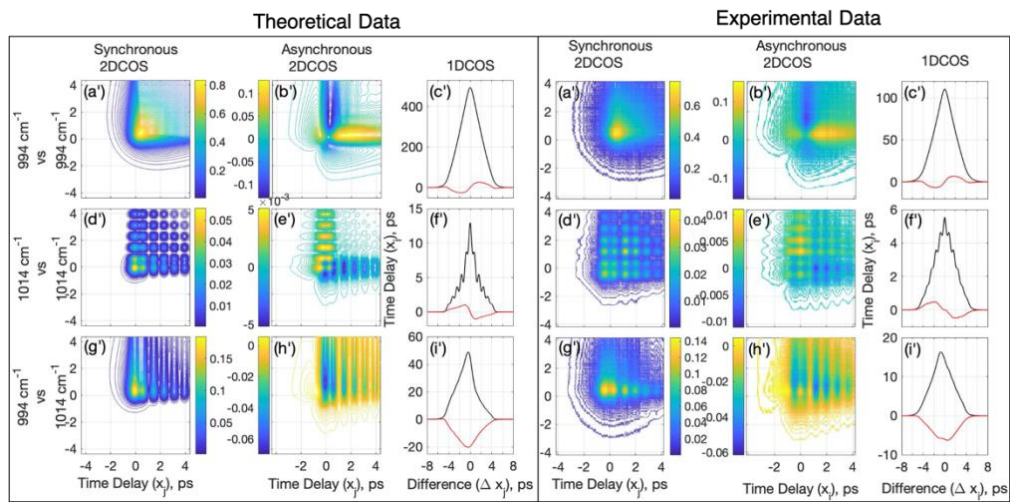

Figure 7: Theoretical (left set) and experimental (right set) synchronous 2DCOS (first column), asynchronous 2DCOS (middle column) and synchronous (black) and asynchronous 1DCOS with autocorrelations of pyridine intensities at 994 cm$^{-1}$ (first row) and at 1014 cm$^{-1}$ (middle row) and their cross-correlations in last row.

The first three columns (left set) in Fig. 7(a -i) depict the theoretical simulations of 2D and 1D correlation results for pyridine intensities at 994 cm$^{-1}$ and 1014 cm$^{-1}$. The first and the middle row are dedicated to the 1D/2D autocorrelations for water/water and benzene/benzene, respectively. The last row shows the results for their cross-correlations. The experimental results are plotted in the same manner in the last three columns of Fig.7 (a'-i'). The synchronous 2D autocorrelations of the peak at 994 cm$^{-1}$ in Fig. 7(a) shows an off-center peak that is getting wider in the first (i.e., positive time delay) quadrant. The synchronous 2D autocorrelation of the peak at 1014 cm$^{-1}$ is symmetrically centered at zero delay of the probe pulse delay (see Fig. 7(d)), along with which beating can be observed in the positive quadrant. Asynchronous 2D autocorrelation of the peak at 994 cm$^{-1}$ (see Fig. 7(b)) depicts a positive peak whose maxima is shifted at 2 ps delay of the probe. However, the most exciting result is seen in the 2D cross-correlations of peaks at 994 cm$^{-1}$ and 1014 cm$^{-1}$ (see Fig 7(h)). When visualizing the time delay of the peak at 994 cm$^{-1}$ on the vertical axis and time delay of the peak at 1014 cm$^{-1}$ on the horizontal axis, it clarifies that there is no change in the horizontal direction with a change in probe width. However, the peak on the vertical axis shows an alteration. Diagonally summed results of synchronous and asynchronous 2D correlations results reproduced using Eq. (8) and (9) are depicted with black and red curves in the last columns of theoretical and experimental data. A uniform beating pattern is observed in the synchronous 1D autocorrelation of the peak at 1014 cm$^{-1}$ (see the black curve in Fig. 7(f)). This black curve shows the autocorrelations of the quantum beating with itself.

## Conclusions

In this work, we present an application of the 1D/2DCOS analyses to successfully distinguish between non-resonant and resonant contributions in the overall CARS signal. Differentiation of these processes' contribution has been a challenge in the coherent Raman spectroscopy community; hence, this tool could be employed to effectively segregate the NR background using the CARS deferral dependence buildup delay time on its spectral probe width. One- and two-dimensional intensity-intensity correlation functions in terms of the new probe pulse delay variable and new probe pulse linewidth perturbation parameter are mapped to observe transient features of coherent Raman signal buildup. In conjunction with a one-dimensional correlation function shown to be directly derivable from the diagonal directional sum (i.e., diagonal projection) of the synchronous plot, we introduce here a new definition for a pair of 1D correlation functions corresponding to synchronous and asynchronous 2DCOS data. This analysis can be used in the coherent Raman community as an indispensable tool in bringing out obscured spectral features and understanding the interdependence of variables studied via spectral responses together with the other external experimental parameters. Theoretically, simulated non-resonant and resonant responses using an all Gaussian model are analyzed through 1D/2DCOS. The theoretical models agree with experimental results obtained with water representing a non-resonant process, benzene and pyridine representing the resonant process. As in conclusions, the present intensity-intensity correlation analytical tool may hold great potential in resolving and visualizing resonant versus non-resonant four wave mixing processes for quantitative label-free species-specific nonlinear spectroscopy and microscopy.

## Disclosures

The authors declare no conflicts of interest